\documentclass{article}
\usepackage{graphicx}
\def\no{\noindent}

\def\bc{\begin{center}}
\def\ec{\end{center}}

\def\beq{\begin{equation}}
\def\eeq{\end{equation}}
\def\no{\noindent}

\def\u{\uparrow }
\def\d{\downarrow }
\def\bt{{\bar t}}

\begin{document}
\title{
Spin-1/2 fermions: crossover from weak to strong attractive 
interaction}

\author{K. Ziegler\\
Institut f\"ur Physik, Universit\"at Augsburg, Germany\\
e-mail: ziegler@physik.uni-augsburg.de
}

\maketitle

Abstract: 

The formation and dissociation of bosonic molecules in an optical
lattice, formed by spin-1/2 fermionic atoms, is considered in the 
presence of an attractive nearest-neighbor interaction. A mean-field
approximation reveals three different phases
at zero temperature: an empty, a condensate and a Mott-insulating
phase. The density of fermionic atoms and the density of bosonic 
molecules indicate a characteristic behavior with respect to the
interaction strength that distinguishes between a dilute and a dense
regime. In particular, the attractive interaction favors the formation of
molecules in the dilute regime and the dissociation of atoms in the
dense regime.


\section{Introduction}

Feshbach resonances provide a powerful tool to vary the interaction
strength of atomic gases \cite{kleppner04}. 
Physical effects, like the formation
of molecules, Bose-Einstein condensation or Cooper pairing in fermionic
systems depend strongly on the strength of the interparticle interaction 
\cite{strecker03,salomon03,zwierlein03,grimm03,jin03}.
A model for spin-1/2 fermions with attractive nearest-neighbor interaction
is studied to describe the formation and dissociation of bosonic molecules
in a grand-canonical gas of fermionic atoms. 
The model depends on three tunable parameters, the fermionic tunneling rate 
$\bt$, the fugacity $\zeta$ of the grand-canonical gas, and the molecular 
tunneling rate $J$. $J$ also controls the attractive nearest-neighbor 
interaction of the model. 

A difference in comparison with previous works 
\cite{holland01,chiofalo02,ohashi02,holland04} is that we study the $T=0$
properties of a Fermi gas in an optical lattice. A consequence of the latter
is that a Mott-insulating state can be identified, whereas $T=0$ avoids
thermal fluctuations of the system. In this case a mean-field
approximation should give reasonable results, provided that quantum 
fluctuation are not too strong.

A qualitative picture of the $T=0$ phase diagram can be given by 
an estimate of the energies. There is a single-particle potential 
due to the chemical potential
of the fermionic atoms. It is assumed that the chemical potential of the
molecules is the same as that of the constituting fermionic atoms, i.e. the
effective chemical potential of a molecule is twice as that of
a single fermionic atom. If this potential is sufficiently negative 
the optical lattice is empty, since particles cannot overcome the potential
barrier.
The kinetic energy, however, represented by the tunneling rates $\bt$ and
$J$, competes with the chemical potential and favors the creation of a Fermi
gas and/or a molecular Bose gas. The interaction plays a crucial role in this
regime: In addition to the formation of (local) bosonic
molecules, the attractive interaction between the fermionic atoms may lead
to the formation of Cooper pairs, which can be considered as ``non-local 
molecules''.
For sufficiently large chemical potential the optical lattice will be filled
with two atoms per lattice site, which is a Mott-insulating state. This Mott
state differs from the Mott insulator of a half-filled Fermi lattice
gas with {\it repulsive} interaction (Hubbard model \cite{fulde}), where the 
interaction creates a ground state with one fermion per lattice site.
In other words, the Mott insulator in the system with attractive interaction
is a special kind of insulator that is created by the Pauli exclusion, and
not by a finite repulsive interaction, in contrast to the
Hubbard model at half filling.

A mean-field theory is applied to
discuss a Mott insulator and a condensed phase. Depending on the tunable
parameters of the model, the density of dissociated atoms 
and the density of molecules 
in the condensed phase are investigated. It will be shown that the
density of dissociated atoms increases
with the total number of particles, reaches a maximum and decreases, whereas
the density of molecules is a monotoneously increasing function.

\section{Model: Hamiltonian and Functional Integral}

A gas of spin-1/2 fermions in an optical lattice is considered. Using
fermionic creation (annihilation) operators $c^\dagger_{r\sigma}$
($c_{r\sigma}$) for spin $\sigma$ and at site $r$ (a minimum in the
optical lattice), the Hamiltonian of the Fermi gas is 
\beq
H=-\bt\sum_{<r,r'>}\sum_{\sigma=\u,\d}c^\dagger_{r\sigma}c_{r'\sigma}
-J\sum_{<r,r'>}c^\dagger_{r\u}c_{r'\u}c^\dagger_{r\d}c_{r'\d}
-\mu\sum_{r}\sum_{\sigma=\u,\d}c^\dagger_{r\sigma}c_{r\sigma}.
\label{hamilton}
\eeq
$<r,r'>$ refers to nearest-neighbor sites $r$ and $r'$. A chemical
potential term for molecules
\[
-\nu\sum_r c^\dagger_{r\u}c_{r\u}c^\dagger_{r\d}c_{r\d}
\]
has been neglected because it leads only to a shift of $J$ in terms
of the subsequent mean-field calculation.
Similar Hamiltonians were considered in a number of papers 
\cite{holland01,chiofalo02,ohashi02}. 
The first term describes tunneling of individual fermions in the
optical lattice with rate $\bt$, the second term tunneling of local
fermion pairs (i.e. bosonic molecules) between nearest-neighbor 
sites with rate $J$.
It should be noticed that the latter is also responsible for an
attractive interaction between fermions with different spin
($J>0$). However, there is no local (diagonal) interaction,
all the fermionic interaction is carried by the nearest-neighbor 
term. The bosonic molecules experience a local repulsive (hard-core)
interaction, since their constituent atoms are fermions and obey the
Pauli exclusion. 
For the limiting case $\bt=0$ only bosonic molecules can
appear, for $J=0$ only non-interacting fermionic atoms. The chemical
potential $\mu$ controls the number of particles in
a grand-canonical ensemble. The latter is given by the partition function
\[
Z=Tr e^{-\beta H}.
\]
Space-time correlations of fermions are described by the Green's function
\[
G_{r,t,\sigma ;r',0,\sigma'}= {1\over Z} Tr\Big[ e^{-(\beta-t) H}
c_{r,\sigma}e^{-tH}c^\dagger_{r',\sigma'}\Big].
\]
The partition function can also be written in terms of a Grassmann integral
\cite{negele,ziegler02} as
\[
Z=\int e^{-S}{\cal D}[\psi],
\]
where the action $S$ for spin-1/2 fermions with attractive interaction,
related to the Hamiltonian in Eq. (\ref{hamilton}), reads
\[
S=\sum_x(\psi_{x,1}{\bar\psi}_{x,1}
+\psi_{x,2}{\bar\psi}_{x,2})
-\sum_{r,r',t}(\zeta\delta_{r,r'}+\bt w_{r,r'})
(\psi_{r,t,1}{\bar\psi}_{r',t+1,1}
+\psi_{r,t,2}{\bar\psi}_{r',t+1,2})
\]
\beq
-J\sum_{r,r',t}\psi_{r,t,1}{\bar\psi}_{r',t+1,1}\psi_{r,t,2}
{\bar\psi}_{r',t+1,2}
\label{action0}
\eeq
with space-time coordinates $x=(r,t)$.
$w_{r,r'}$ is $1/2d$ for nearest-neighbor sites on a $d$-dimensional
cubic optical lattice and zero otherwise, and $\zeta=e^\mu$ is the fugacity.
The Green's function then reads
\[
G_{x,\sigma ;x',\sigma'}=
\langle  \psi_{x,\sigma}{\bar\psi}_{x',\sigma'}\rangle 
\equiv\int \psi_{x,\sigma}{\bar\psi}_{x',\sigma'} e^{-S}{\cal D}[\psi ]/Z.
\]

After renaming ${\bar\psi}$ by a time shift $t\to t-1$
\beq
{\bar\psi}_{r,t,j}\to{\bar\psi}_{r,t-1,j}\ \ \ (j=1,2)
\label{shift}
\eeq
the action reads
\[
S'=\sum_x(\psi_{x,1}\partial_t^T{\bar\psi}_{x,1}
+\psi_{x,2}\partial_t^T{\bar\psi}_{x,2})
-\sum_{r,r',t}(\zeta\delta_{r,r'}+\bt w_{r,r'})
(\psi_{r,t,1}{\bar\psi}_{r',t,1}
+\psi_{r,t,2}{\bar\psi}_{r',t,2})
\]
\[
-J\sum_{r,r',t}\psi_{r,t,1}{\bar\psi}_{r',t,1}\psi_{r,t,2}
{\bar\psi}_{r',t,2}.
\]
The quartic interaction term is now diagonal with respect to time.
It can be decoupled by two complex Hubbard-Stratonovich
fields $\phi_x$, $\chi_x$ \cite{ziegler02}. The linear combination
$i\phi+\chi$ couples to fermions like
\beq
(i\phi_{x}+\chi_{x})\psi_{x,1}\psi_{x,2}+h.c..
\label{eff}
\eeq
This coupling is similar to the atom-molecule coupling discussed 
in Refs. \cite{holland04,petrov03}. There is a difference, however,
by the fact that we started from the fermionic model and derived
an effective electron-boson model, whereas the other authors started 
from an electron-boson model and derived the effective fermion
model. In both cases an effective coupling constants can be evaluated
from the original model, an effective fermion-fermion coupling
in Ref. \cite{holland04} and an effective fermion-boson coupling in
our model. Although this is an interesting problem it will not be
pursued in this paper. Instead, the fermions are integrated out in $Z$, 
since they appear only in a quadratic form in the action. This step
provides an effective model only for bosons: the integration
gives a fermion determinant and the resulting
fuctional $Z$ depends only on the complex fields $\phi$ 
and $\chi$ with the effective action
\beq
S_{eff}=(\phi,v^{-1}\phi)+{1\over 2J}(\chi,\chi)
-{1\over 2}\log(\det A)
\label{seff}
\eeq
with the antisymmetric space-time matrix
\[
A=\pmatrix{
0 & i\phi+\chi & \zeta+\bt w-\partial_t^T & 0 \cr
-i\phi-\chi & 0 & 0 & \zeta+\bt w - \partial_t^T \cr
-\zeta-\bt w +\partial_t & 0 & 0 & i{\bar\phi}+{\bar\chi} \cr
0 & -\zeta-\bt w +\partial_t & -i{\bar\phi}-{\bar\chi} & 0 \cr
}
\]
with $v^{-1}=(w+2{\bf 1})^{-1}/J$.

$\phi$ corresponds to the
conventional BCS field in the case of a local interaction.
The additional field $\chi$ is necessary for the  nearest-neighbor 
interaction to ensure that the Hubbard-Stratonovich decoupling is 
well-defined \cite{ziegler02}. It will
be discussed subsequently that the mean-field approximation
yields a linear relationship between these two fields. 

\subsection{Atomic and Molecular Densities}

The densities of atoms and molecules can be measured as expectation 
values of the Grassmann fields. For this purpose the original fields are
used, i.e., the fields before the time shift in Eq. (\ref{shift}), to write 
\[
\langle ... \rangle =\int ... e^{-S}{\cal D}[\psi ]/Z.
\]
The action $S$ of Eq. (\ref{action0}) can be expanded in $Z$ around 
\[
S_0=\sum_x(\psi_{x,1}{\bar\psi}_{x,1}+\psi_{x,2}{\bar\psi}_{x,2})
\]
to obtain
\[
Z=\sum_{l\ge0}{1\over l!}\int e^{-S_0}(S_0-S)^l{\cal D}[\psi ].
\]
This expansion can be viewed as an expansion in terms of
world lines in a space-time lattice. There are two types of
world lines, individual fermion lines with spin $\sigma$ and
molecular world lines, as shown in Fig. 1. For a given point
in space and time 
$x$ the contribution from the polynomial $(S_0-S)^l$ is
either a factor 1 (i.e. no contribution from $(S_0-S)^l$ at this point),
a factor $\psi_{x,\sigma}{\bar\psi}_{x,\sigma}$, or a factor
$\psi_{x,1}{\bar\psi}_{x,1}\psi_{x,2}{\bar\psi}_{x,2}$. 
To measure the probability of the appearence of these factors in $Z$,
an expectation value with respect to the Grassmann field can be introduced.
In particular, the density of dissociated atoms reads
\[
n_{f,x}=-\langle\psi_{x,1}{\bar\psi}_{x,1}
+\psi_{x,2}{\bar\psi}_{x,2}+2\psi_{x,1}{\bar\psi}_{x,1}
\psi_{x,2}{\bar\psi}_{x,2}\rangle
\]
and the density of molecules
\[
n_{m,x}=\langle 1+\psi_{x,1}{\bar\psi}_{x,1}
+\psi_{x,2}{\bar\psi}_{x,2}+\psi_{x,1}{\bar\psi}_{x,1}
\psi_{x,2}{\bar\psi}_{x,2}\rangle .
\]
The truncated expectation value
\[
C_{12}=\langle\psi_{x,1}{\bar\psi}_{x,1}
\psi_{x,2}{\bar\psi}_{x,2}\rangle-
\langle\psi_{x,1}{\bar\psi}_{x,1}\rangle\langle
\psi_{x,2}{\bar\psi}_{x,2}\rangle
\]
gives
\beq
n_{f,x}=-\langle 1+\psi_{x,1}{\bar\psi}_{x,1}\rangle
\langle\psi_{x,2}{\bar\psi}_{x,2}\rangle
-\langle1+\psi_{x,2}{\bar\psi}_{x,2}\rangle
\langle\psi_{x,1}{\bar\psi}_{x,1}\rangle -2C_{12}
\label{denf}
\eeq
and
\beq
n_{m,x}=\langle 1+\psi_{x,1}{\bar\psi}_{x,1}\rangle
\langle 1+\psi_{x,2}{\bar\psi}_{x,2}\rangle + C_{12}.
\label{denm}
\eeq
It will be seen below that $C_{12}$ vanishes in mean-field
approximation.

In the Mott-insulating phase a fully occupied lattice is expected, i.e.,
two fermionic atoms (= one molecule) per site, as the only commensurate
state, since there is no repulsive interaction which could maintain a
commensurate state. Any other groundstate of the system is a condensate,
except for the empty lattice. The remaining question is how the condensate
varies with the total number of atoms and molecules in the system.
It is obvious that the total density of particles increases monotoneously
with the fugacity. However, it is less clear how the formation of
molecules or the dissociation of atoms is affected by an increasing
fugacity. Moreover, a naive picture suggests that the density of molecules
increases monotoneously with an increasing attractive coupling of the
fermionic atoms $J$. It will be shown 
in terms of a mean-field approximation that this is not the case.

\section{The Saddle-point Equation}

The saddle-point condition for the effective action in Eq. (\ref{seff}) is
\beq
\delta S_{eff}=0.
\label{spe0}
\eeq
This gives a nonlinear difference equation. A simple ansatz for
its solution is given by uniform fields $\phi$ and $\chi$ (mean-field
approximation). Then Eq. (\ref{spe0}) reads
\beq
\cases{
\phi=3JG(\phi-i\chi) \cr
\chi=-i2JG(\phi-i\chi)\cr
}
\label{spe}
\eeq
with the integral
\beq
G=\int_{-1}^1{\rho(x)\over\sqrt{
(-i\chi+\phi)^2({-i\bar\chi}+{\bar\phi})^2
+2(s^2+1)(-i\chi+\phi)(-i{\bar\chi}+{\bar\phi})
+(s^2-1)^2
}} 
dx
\label{int1}
\eeq
and $s=\zeta+\bt x$. $\rho(x)$ is the density of states
for the nearest-neighbor tunneling term $w$.

There is a trivial solution $\phi=\chi=0$ and a nontrivial solution
\beq
G=1/J,\ \ \ \chi=-2i\phi/3.
\label{bcs}
\eeq
This is the same mean-field equation as in the BCS theory, if $J$ is
considered as the coupling constant of the fermions. 

The solution in Eq. (\ref{bcs}) implies a non-negative value for
\[
(-i\chi+\phi)(-i{\bar\chi}+{\bar\phi})=\phi{\bar\phi}/9\equiv |\phi|^2/9.
\]
Then $G$ in Eq. (\ref{int1}) reads
\beq
G=\int_{-1}^1{\rho(x)\over\sqrt{
|\phi|^4/81+2(s^2+1)|\phi|^2/9+(s^2-1)^2
}} 
dx
\label{spe2}
\eeq
which descreases monotoneously with $|\phi|^2$.
If $(s^2-1)^2$ becomes 0 inside the interval of integration $-1\le x\le 1$
the integral diverges for $|\phi|\to0$. Therefore, there is a non-zero
solution of $\phi$ for any value of $J>0$. On the other hand, if $(s^2-1)^2$
does not become 0 inside the interval of integration, there is a non-zero 
solution only for sufficiently large values of $J$ (cf. Fig. 2). 

The densities can be evaluated within the saddle-point approximation.
A straightforward calculation gives for the expectation values, used
in $n_{f,x}$ and $n_{m,x}$, the expression
\beq
\langle\psi_{x,j}{\bar\psi}_{x,j}\rangle 
=-{1\over 2}+{1\over 2}\int_{-1}^1
{|\phi|^2/9+s^2-1\over\sqrt{
|\phi|^2/81+2(s^2+1)|\phi|^2/9+(s^2-1)^2}} \rho(x)dx
\label{exp1}
\eeq
and
\[
\langle\psi_{x,j}{\bar\psi}_{x,j'}\rangle =0\ \ \ (j\ne j').
\]

\section{Results}

Using the mean-field approximation of the previous section 
the following quantities are evaluated: 
the order parameter $|\phi|^2$ of the condensate,
the $T=0$ phase diagram,
and $n_f(|\phi|^2,\zeta)$, $n_m(|\phi|^2,\zeta)$.

\subsection{$\phi=0$}

In the case of a vanishing order parameter (i.e. $\phi=0$)
only independent fermions are described by the mean-field
approach.  Then the diagonal element in Eq. (\ref{exp1}) is
\[
\langle\psi_{x,j}{\bar\psi}_{x,j}\rangle 
=-\int_{-1}^1\Theta(1-(\zeta+tx)^2)\rho(x)dx.
\]
If $s^2<1$ for the entire interval of integration, i.e., 
\[
-1+\bt <\zeta<1-\bt
\]
both densities vanish: $n_f=n_m=0$. On the other hand, 
if $s^2>1$ for the entire interval of integration, i.e.,
\[
\zeta>1+\bt
\]
one obtains $n_f=0$, $n_m=1$. Thus all lattice sites are occupied by
bosonic molecules. There is also an intermediate regime
for $1-\bt<\zeta<1+\bt$ where $n_f, n_m\ne0$. Details of this
behavior are shown in Figs. 2 and 3.

\subsection{$\phi\ne 0$}

For the beginning the special case $\bt=0$ (molecules cannot dissociate) 
is considered. Then the integrand in Eq. (\ref{exp1}) is
constant with respect to the integration variable $x$ and yields
\[
\langle\psi_{x,j}{\bar\psi}_{x,j}\rangle =-1/2
+{1\over2}
{|\phi|^2/9+\zeta^2-1\over\sqrt{
|\phi|^4/81+2(\zeta^2+1)|\phi|^2/9 +(\zeta^2-1)^2
}}.
\]
The saddle-point solution 
\[
|\phi|^2=9(\sqrt{4\zeta^2+J^2}-1-\zeta^2)
\]
exists for 
\[
\cases{
\sqrt{1-J}<\zeta<\sqrt{1+J} & if $0<J\le1$ \cr
0<\zeta<\sqrt{1+J} & if $J >1$ \cr
}.
\]
The corresponding phase diagram is shown in Fig. 4. This result yields
\[
\langle\psi_{x,j}{\bar\psi}_{x,j}\rangle=-1/2-1/J+\sqrt{\zeta^2/J^2+1/4}.
\]
In the general case with dissociated atoms (i.e. for $\bt>0$),
the density of states in $G$ can be approximated by a constant as
$\rho(x)=1/2$. Moreover, for small $J$ the integral in Eq. (\ref{spe2})
can be easily performed, giving the saddle-point solution
of the order parameter
\[
|\phi|^2\sim 4\bt^2 e^{-4\bt /J}\ \ \ \ (1-\bt <\zeta< 1+\bt).
\]

The model parameters $\zeta$, $J$, and $\bt$ control the densities.
However, in a realistic situation the order parameter $\phi$ and the 
densities $n_f,n_m$ will be measured. Therefore, the densities 
for a given value of $|\phi|$ are plotted as functions of the 
fugacity $\zeta$, where increasing of $|\phi|$ means an intereasing
interaction parameter $J$. The result is shown in Figs. 2 and 3:
(1) The density of dissociated atoms $n_f$ has a maximum. This maximum is
at $\zeta=1$ for $\phi=0$ and moves to lower values of $\zeta$ with 
increasing order parameter values. The maximum value of $n_f$ itself is $0.5$
but decreases for $|\phi|>1$.
(2) The density of molecules $n_m$ always increases with $\zeta$. But as a
function of $|\phi|$ it increases (decreases) below (above) $\zeta\approx1$.
This result
indicates that strong interaction favors the formation of molecules in a
dilute system and favors the dissociation of atoms in a dense system. 
The latter can be understood as a special kind of frustration effect because 
there many ways for a pair of fermionic atoms to form a molecule. 
(3) The total density $n_f+n_m$ increases always monotoneously with $\zeta$.

\section{Conclusions}

An attractively interacting Fermi gas in an optical lattice was treated in
mean-field approximation. At zero temperature three different phases were
found: an empty phase, a Mott-insulating phase and a condensed phase. 
The latter is characterized by a non-vanishing order parameter. The
fermions appear in this phase as a mixture of local pairs (molecules) 
and extended pairs (dissociated atoms). The densities of these two types 
of fermionic pairs have a characteristic behavior for the crossover 
from weak to strong attractive interaction. This is indicated by
(1) a maximum of the density of dissociated fermionic atoms and (2) 
support for the formation of molecules (dissociation of atoms) in the 
dilute (dense) system by the attractive interaction.
  
\vskip0.5cm
\no
{\bf Acknowledgement}

\no
The author is grateful to G. Shlyapnikov for useful discussions.
This research was supported in part by the National Science Foundation
under Grant No. PHY99-07949 and by the Deutsche Forschungsgemeinschaft
through SFB 484.



\newpage


\begin{figure} 
\begin{center}
\includegraphics[scale=0.6]{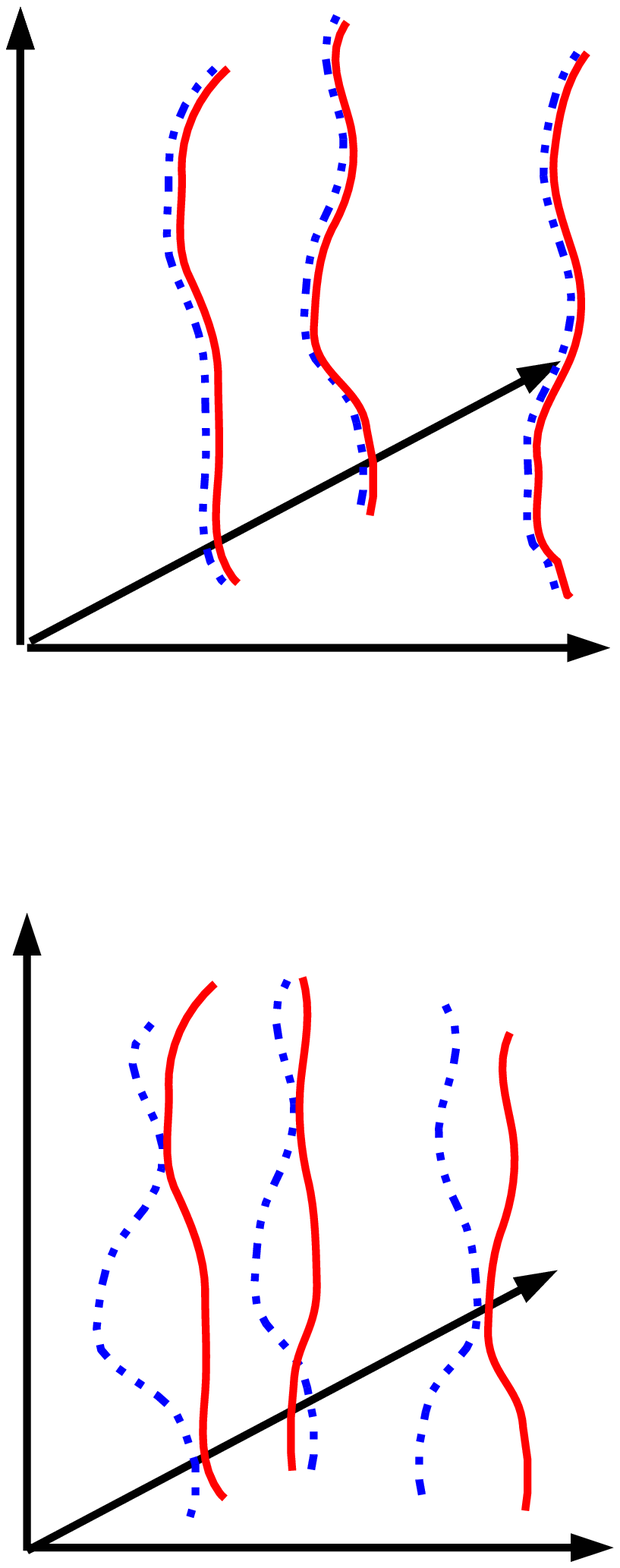}
\end{center}
\caption{Statistics of world lines in a pure system of bosonic molecules (a)
and in a mixture of fermionic atoms and bosonic molecules (b). Full and dashed 
lines distinguish the fermionic spin $\u$ and $\d$.}
\end{figure}

\begin{figure} 
\begin{center}
\includegraphics[scale=0.8]{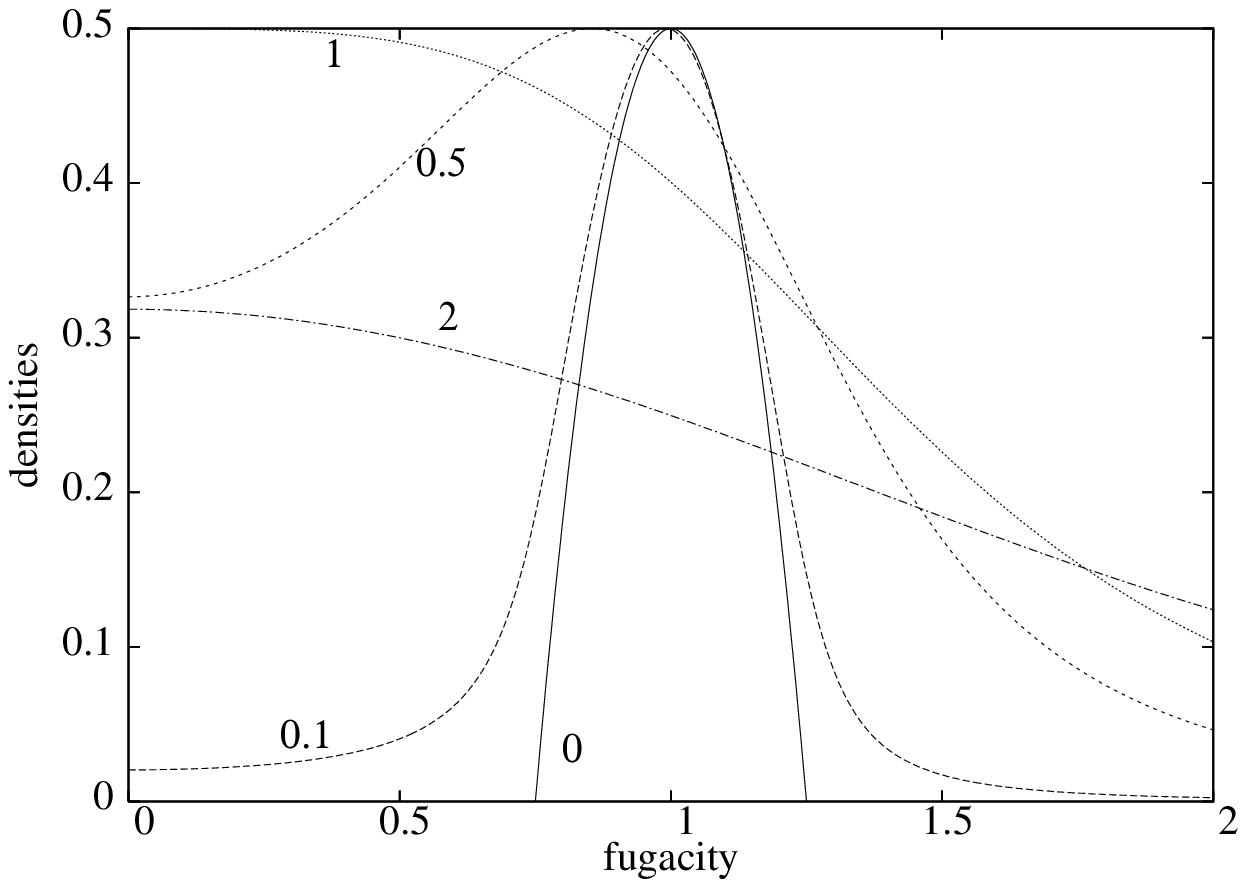}\\
\includegraphics[scale=0.8]{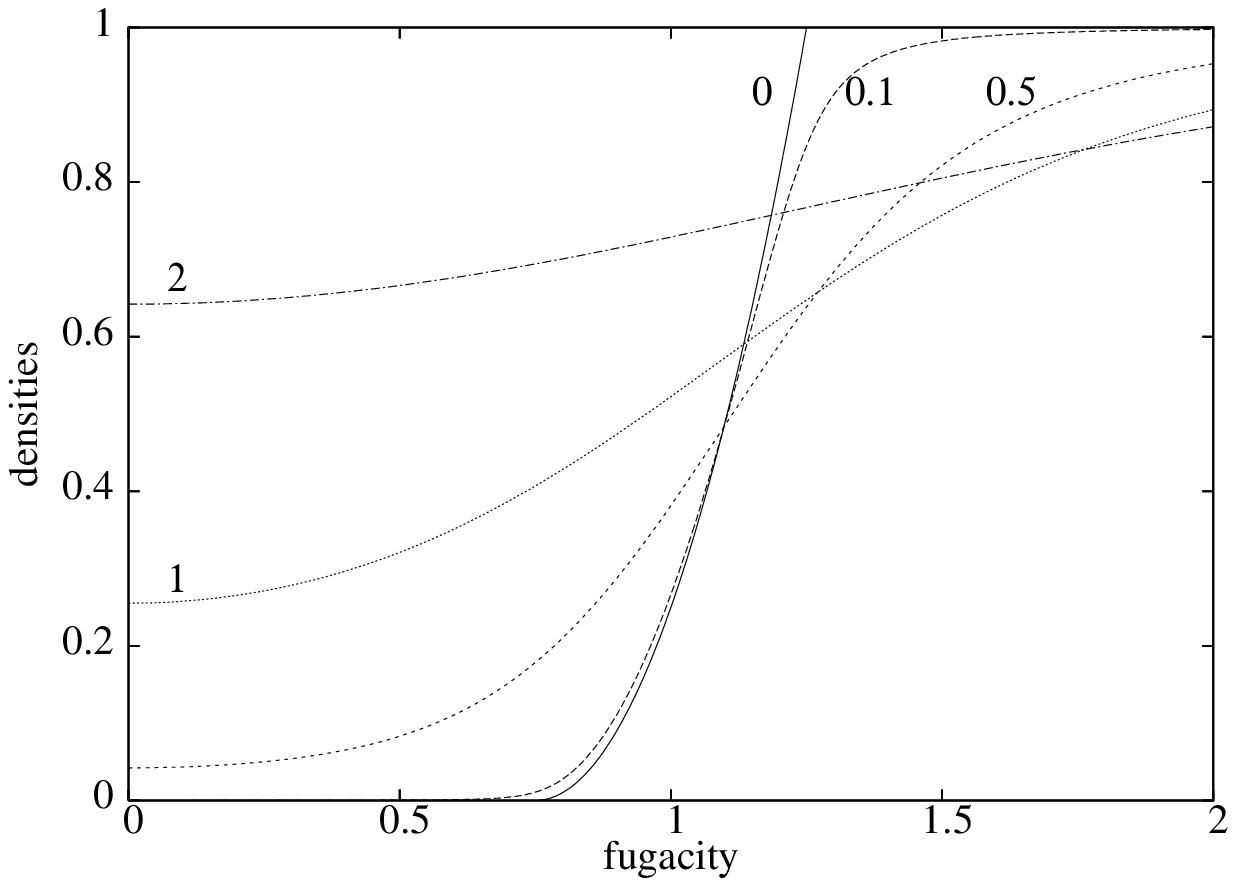}
\end{center}
\caption{The density of individual fermions (dissociated atoms) 
(first figure)
and molecules (second figure) for different order parameter values
$\phi=0$, $|\phi|=0.1$, $|\phi|=0.5$, $|\phi|=1$, and $|\phi|=2$, 
all with atomic tunneling rate $\bt=0.25$.}
\end{figure}
\begin{figure} 
\begin{center}
\includegraphics[scale=0.8]{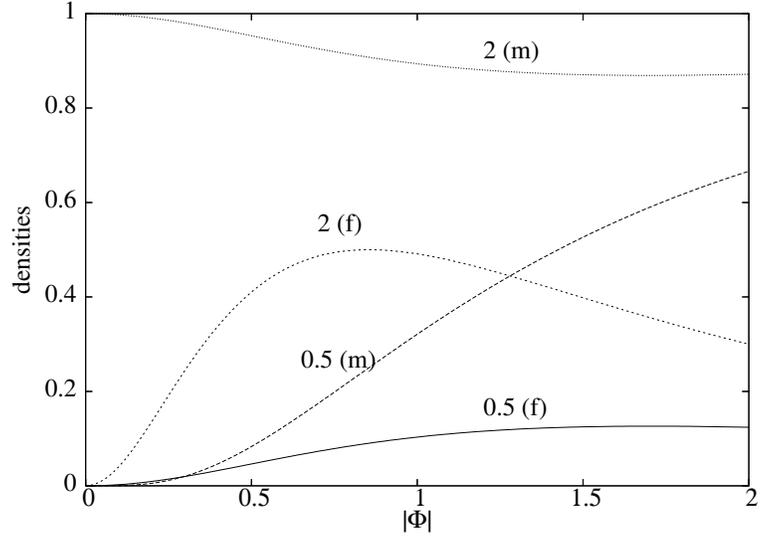}
\end{center}
\caption{The densities as a function of the order parameter.
Numbers refer to the value of the fugacity $\zeta$ and letters to
fermionic atoms (f) and molecules (m).}
\end{figure}
\begin{figure} 
\begin{center}
\includegraphics[scale=0.8]{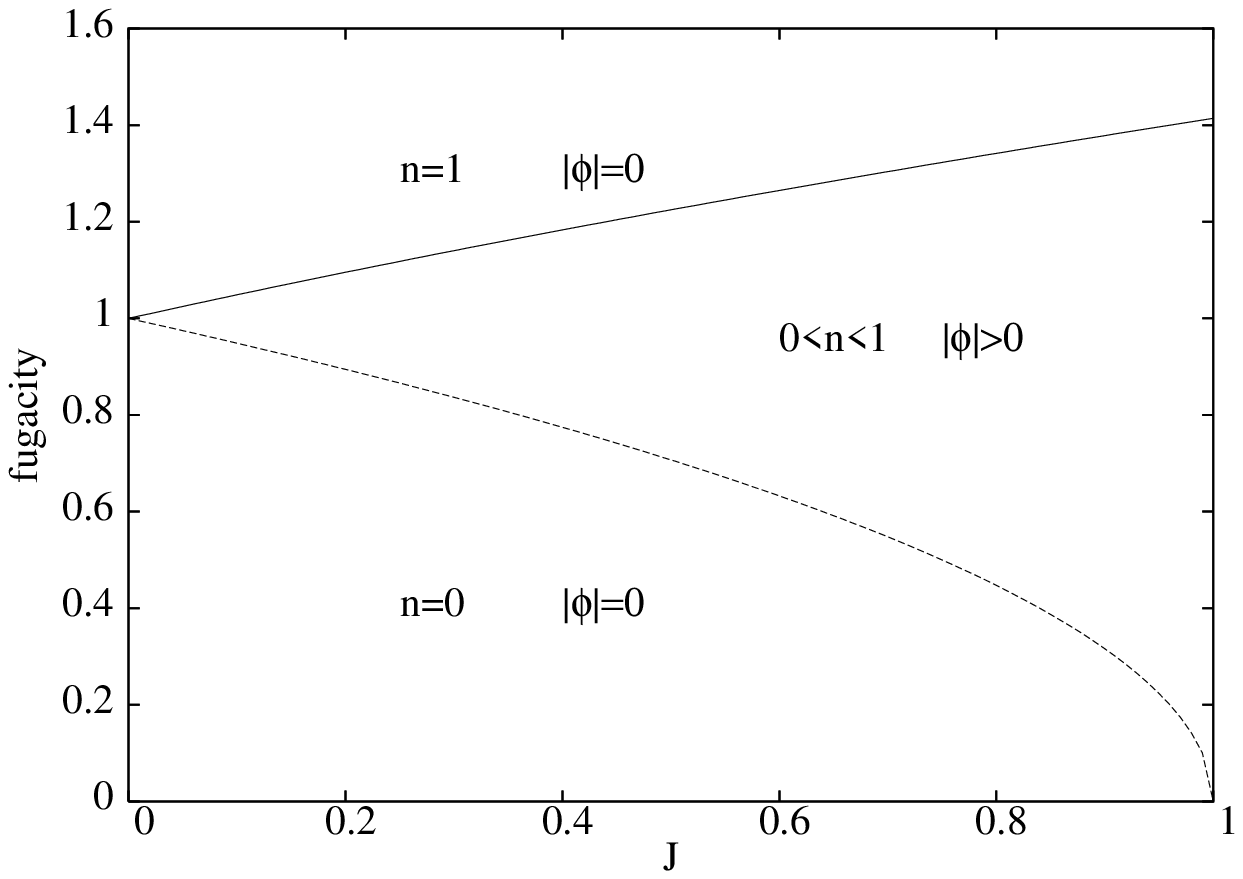}
\end{center}
\caption{$T=0$ phase diagram for atomic tunneling rate $\bt=0$:
fugacity $\zeta$ vs. molecular tunneling rate $J$.
The indicated densities $n$ are molecular densities $n_m$.
The condensed phase with $|\phi|>0$ becomes wider when $\bt>0$.
In particular, the boundaries of this phase at $J=0$ are
$1-\bt$ and $1+\bt$.}
\end{figure}


\begin{thebibliography}{99}

\bibitem{kleppner04}
Kleppner, D., 2004, Physics Today August, 12.

\bibitem{strecker03}
Strecker, K.E. et al., 2003, Phys. Rev. Lett. {\bf 91}, 080406. 

\bibitem{salomon03}
Cubizolles, J. et al., 2003, Phys. Rev. Lett. {\bf 91}, 240401.

\bibitem{zwierlein03}
Zwierlein, M.W. et al., 2003, Phys. Rev. Lett. {\bf 91}, 250401.

\bibitem{grimm03} 
Jochim, S. et al., 2003, Phys. Rev. Lett. {\bf 91}, 240401.

\bibitem{jin03}
Regal, C.A., Greiner, M., and Jin, D.S., 2003, Phys. Rev. Lett. {\bf 92},
083201.

\bibitem{dickerscheid}
Dickerscheid, D.B.M. et al., 2004, Feshbach resonances in an optical lattice,
cond-mat/0409416.
 
\bibitem{holland01}
Holland, M. et al., 2001,
Phys. Rev. Lett. {\bf 87}, 120406.

\bibitem{chiofalo02}
Chiofalo, M.L. et al., 2002, Phys. Rev. Lett. {\bf 88}, 090402.

\bibitem{ohashi02}
Ohashi, Y. and Griffin, A., 2002, Phys. Rev. Lett. {\bf 89}, 130402.

\bibitem{holland04}
Holland, M.J., Menotti, C., and Viverit, L., 2004, cond-mat/0404234.

\bibitem{petrov03}
Petrov, D.S., Salomon, C., and Shlyapnikov, G.V., 2003, cond-mat/0309010.

\bibitem{negele}
Negele, J.W. and Orland, H., 1988, {\sl Quantum Many - Particle Systems}
(New York: Addison - Wesley).

\bibitem{fulde}
Fulde, P., 1993, {\sl Electron Correlations in Molecules and Solids}
(Berlin: Springer - Verlag).

\bibitem{ziegler02}
Ziegler, K., 2002, Journ. Low. Temp. Phys. {\bf 126}, 1431.

\end{thebibliography}
\end{document}